\documentclass[fleqn,twoside]{article}
\usepackage[fleqn]{amsmath}
\usepackage{espcrc2}

\usepackage{epsfig}

\usepackage[figuresright]{rotating}

\title{Measurements of time dependent CP asymmetry in $B \rightarrow
 VV$ decays with BELLE}

\author{Ryosuke Itoh\address[KEK]{Institute of Particle and Nuclear Studies, 
        High Energy Accelerator Research Organization (KEK), \\ 
        1-1 Oho, Tsukuba, Ibaraki 305-0801, Japan }%
\\ Representing the Belle Collaboration
}
       
\begin{document}

\begin{abstract}
A study of CP violation in $B\rightarrow
J/\psi K^*(K_S^0\pi^0)$ decays by 
time dependent angular analysis is discussed.
Status of time independent analyses for other
$B\rightarrow VV$ decays
is also reported. 
The data used for the analyses are taken with the Belle detector at
KEK. 

\end{abstract}

\maketitle

\section{Introduction}

There are three helicity states in $B \rightarrow
VV$ decays. Although one of the states is a pure CP even state, 
CP even and odd states are mixed in other states and the
observed CP asymmetry is diluted. 
By studying decay angles of final state particles,
it is possible to project out each CP
state in a statistical way. 
A theoretical expression of 
differential time-angular decay rate of $B \rightarrow VV$
decays consists of components corresponding to three helicity 
states and also to
the interference between them. Each term is
expressed as a product of an angular term and an amplitude term. The
amplitude term contains CP violating phase(s) as a function of
$\Delta t$, the decay time difference between two $B$ mesons from
an $\Upsilon(4S)$ decay. A fit of the expression
for the measured decay angles and $\Delta t$ gives
the determination of the CP violating phase(s). 
The interference terms are also rich sources of various interesting physics 
such as the $\cos 2\phi_1$ measurement in 
$B^0\rightarrow J/\psi K^{*0}(K_S^0\pi^0)$ decays, and the simultaneous 
determination of $r$ and $\sin(2\phi_1+\phi_3)$ in $B^0\rightarrow
D^{*-}\rho^{+}$ decays where $r$ is the ratio of doubly-Cabibbo-suppressed
mode to Cabibbo favored mode contributing to the decay.

In this talk, a study of time dependent angular analysis for
$B\rightarrow J/\psi K^*(K_S^0\pi^0)$ is
discussed. Also reported is
status of angular analyses for $B^0 \rightarrow
D^{*+}D^{*-}$ and $B^0 \rightarrow D^{*-}\rho^+$.
The data used in the analyses are taken with the Belle detector\cite{Belle} in
the KEKB accelerator\cite{KEKB}. 
The KEKB accelerator consists of two separate rings for 
electrons and positrons. The
energies are set at 3.5 GeV for positrons and 8.0 GeV for
electrons, respectively. These two beams are collided at the interaction point 
to produce $\Upsilon(4S)$ in motion for the measurement of $\Delta t$.
The peak luminosity of the
machine has reached  $7.4 \times 10^{33} cm^{-2}sec^{-1}$ which is the
current world record. The total integrated luminosity up to now
is $89.1 fb^{-1}$. 
The Belle detector is placed at the interaction
point of the KEKB accelerator. It is a general-purpose
detector with a wide angle coverage featuring a particle identification
using the Aerogel Cherenkov Counter assembly. 
The data sample used in the analyses corresponds to an integrated
luminosity of 78 fb$^{-1}$ taken on the $\Upsilon(4S)$ resonance.

\section{Time dependent angular analysis for 
$B^0 \rightarrow J/\psi K^*(K_S^0\pi^0)$\cite{belle-conf-0212}}

Candidate $B^0$ mesons are reconstructed by selecting
events with a $J/\psi$ identified from a pair of oppositely-charged leptons
and a $K^*$ from 
a pair of $K_S^0$ and $\pi^0$ candidates. 
The beam-constrained mass ($M_{bc}$), 
which is the invariant mass of a reconstructed
$J/\psi$ and $K^*$ calculated taking the energy to be the beam
energy, is required to be
in the range $5.27-5.29~{\rm GeV}/c^2$ and the energy difference
between $B$ candidate and the beam energy
($\Delta E$) to satisfy $-50~{\rm MeV} <
\Delta E < 30~{\rm MeV}$.
To eliminate slow $\pi^0$ backgrounds,
the angle of the kaon with respect
to the $K^*$ direction in the  $K^*$ rest frame, $\cos \theta_{K^*}$, 
is required to satisfy ${\rm cos}\theta_{K^*}<0.8$. 
When an event contains more than one candidate passing the above requirements,
the combination for which $\Delta E$ is closest to zero is selected.
The number of events remaining after the selection is 104.

The time dependent angular distribution of the decay is described 
as\cite{angleref}:
\begin{gather}
\frac{d^4\Gamma(\theta_{tr},\phi_{tr},\theta_{K^*},\Delta t)}
{d\cos\theta_{tr} d\phi_{tr} d\cos\theta_{K^*} d\Delta t} \nonumber \\
~~~ =  \frac{e^{-|\Delta t|/\tau_{B^0}}}{2\tau_{B^0}} \sum_{i=1}^{6} 
g_i (\theta_{tr},\phi_{tr},\theta_{K^*}) a_i(\Delta t)
\label{tadf}
\end{gather}
where $g_i$ are the angular terms and $a_i$ are amplitude terms.
$\tau_B$ is the lifetime of a $B^0$ meson.
The $g_i$ are expressed as:
\begin{eqnarray}
g_1 & = & 2\cos^2\theta_{K^*}(1-\sin^2\theta_{tr}\cos^2\phi_{tr}) \\
g_2 & = & \sin^2\theta_{K^*}(1-\sin^2\theta_{tr}\sin^2\phi_{tr}) \\
g_3 & = & \sin^2\theta_{K^*}sin^2\theta_{tr} \\
g_4 & = & \frac{-1}{\sqrt{2}}\sin
2\theta_{K^*}\sin^2\theta_{tr}\sin2\phi_{tr} \\
g_5 & = & \sin^2\theta_{K^*}\sin2\theta_{tr}\sin\phi_{tr} \\
g_6 & = & \frac{1}{\sqrt{2}}\sin
2\theta_{K^*}\sin2\theta_{tr}\cos\phi_{tr} 
\end{eqnarray}
where $\theta_{tr}$, $\phi_{tr}$ and $\theta_{K^*}$ are the decay
angles defined in the transversity basis\cite{transversity}.
The $a_i$ are
\begin{eqnarray}
a_1 & = & |A_0|^2(1+\eta\sin2\phi_1 \sin\Delta m\Delta t) \\
a_2 & = & |A_{\parallel}|^2(1+\eta\sin2\phi_1 \sin\Delta m\Delta t) \\
a_3 & = & |A_{\perp}|^2(1-\eta\sin2\phi_1 \sin\Delta m\Delta t) \\
a_4 & = & Re(A_{\parallel}^*A_0)(1+\eta \sin2\phi_1 \sin\Delta
m\Delta t) \\
a_5 & = & \eta Im(A_{\parallel}^*A_{\perp})\cos \Delta m\Delta t -
\nonumber \\
& & 
\eta Re(A_{\parallel}^*A_{\perp})\cos 2\phi_1 \sin \Delta m \Delta t
\\
a_6 & = & \eta Im(A_0^*A_{\perp})\cos \Delta m\Delta t - \nonumber \\
& & \eta Re(A_0^*A_{\perp})\cos 2\phi_1 \sin \Delta m \Delta t 
\end{eqnarray}
where $\Delta m$ is the $B^0-\overline{B^0}$ mixing parameter.
$\eta$ is +1 for $B^0$ while -1 for $\overline{B^0}$.
$A_0$, $A_{\parallel}$ and $A_{\perp}$ are the complex decay
amplitudes for three helicity states in the transversity basis. 
Two CP violation parameters, $\sin2\phi_1$ and $\cos2\phi_1$,
 appear in the formula.
The values of these parameters are determined by
fitting the formula to the measured angles and $\Delta t$,
taking into account the detection efficiency and background.
The fit is done using an unbinned maximum likelihood method. The
probability density function for an event is defined as
\begin{small}
\begin{gather}
PDF =
f_{sig}(M_{bc})
\epsilon(\theta_{tr},\phi_{tr},\theta_{K^*}) \nonumber \\
~~~~ \times \frac{d^4\Gamma(\cos\theta_{tr},\phi_{tr},\cos\theta_{K^*},\Delta t)}
{d\cos\theta_{tr} d\phi_{tr} d\cos\theta_{K^*} d\Delta t}  \nonumber \\
~~ + \frac{e^{-|\Delta t|/\tau_{B^0}}}{2\tau_{B^0}} \nonumber \\ 
~~~~ \times \{\sum_i f_{cf}^i(M_{bc})
              ADF_{cf}(\theta_{tr},\phi_{tr},\theta_{K^*})  \nonumber \\
~~~~~~~~~~ + f_{nr}(M_{bc}) ADF_{nr}(\theta_{tr},\phi_{tr},\theta_{K^*}) \}
\nonumber \\
~~ + \delta(\Delta t) f_{combi}(M_{bc}) 
ADF_{combi}(\theta_{tr},\phi_{tr},\theta_{K^*}). 
\end{gather}
\end{small}
$f_{sig}$, $f_{cf}$, $f_{nr}$ and $f_{combi}$ are fractions of the 
signal, cross feed, non-resonant production, and combinatorial
background as a function of $M_{bc}$, respectively, while
$ADF_{cf}$, $ADF_{nr}$ and $ADF_{combi}$ are corresponding 
angular shape functions. $\epsilon$ is a three
dimensional detection efficiency function for the signal. 
The determinations of 
these functions and the decay amplitudes are described 
in ref.\cite{angana}.
Phases in the amplitudes are chosen so as to conserve $s$-quark
helicity\cite{mahiko}. The $\tau_B$ and $\Delta m$ are fixed at the PDG
values in the fit.
The flavor tagging procedure\cite{Hadronic} gives
the flavor of tag-side $B$ meson $q$ and the probability of wrong tag
$w$. To account for the effect of the wrong tagging in the data, 
$\eta$ is replaced with $-q(1-2w)$ in the PDF.

\begin{figure}
\centerline{\mbox{\psfig{figure=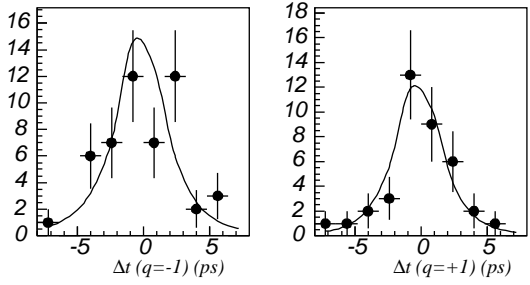,width=7cm}}}
\caption{$\Delta t$ distributions for samples tagged as $q=+1$ and
$-1$. The lines show the predictions by the PDF with determined CP
parameter values.}
\label{deltat}
\end{figure}

Each term in the PDF is then convolved with appropriate resolution
functions\cite{newcp} separately for the signal, the backgrounds with 
a $B^0$ lifetime 
and the combinatorial background with a $\delta(\Delta t)$ function shape.
The parameters in the resolution functions are calculated event by event.
From the fit to the data, we obtain
$	\sin 2\phi_1  =  0.13 \pm 0.51 \pm 0.06, $ and 
$	\cos 2\phi_1  =  1.40 \pm 1.28 \pm 0.19$.
The systematic errors include uncertainties in the resolution
parameters, wrong tagging fractions, decay amplitudes, background
fractions and shapes, and others.
The distributions of $\Delta t$ measured 
for the samples tagged as $q=+1$ and $-1$ 
are shown in Figure~\ref{deltat} together with the predictions by the PDF 
with obtained CP parameter values.

\section{Time independent angular analyses}

The time dependent angular analysis is applicable for other
$B\rightarrow VV$ decays such as $B^0 \rightarrow D^{*+}D^{*-}$, 
$D^{-*}\rho^+$ and 
$\rho^+\rho^-$\footnote{The angular analysis
for $B^+ \rightarrow \rho^+\rho^0$ is covered in other talk\cite{rhorho}.}
for the determination of CP violating phases including
those other than $\phi_1$.
Time dependent analyses for these modes are still
in preparation and the status of time independent(integrated)
analyses is reported here.
The measurement of the decay amplitudes
by the time independent angular analysis is essential for the
determinations of the CP violating phases. 

\subsection{$B^0 \rightarrow D^{*+}D^{*-}$}
The reconstruction of a $D^*$ is performed using a $D\pi$ decay with
a subsequent $D$ decay into $K\pi, K\pi\pi, K\pi\pi\pi, KK$ and $KK\pi$.
A $D$ meson is identified when the invariant mass of decay particles
is within 3 to 6 $\sigma$ from the nominal $D$ mass depending on the
decay mode. A $D^*$ is reconstructed by calculating the mass difference
$M(D\pi)-M(D)$ and those satisfy the difference within
$3.0MeV(D^0)$ or $2.25MeV(D^+)$ are identified as $D^*$. The candidate 
$B$ mesons are selected by applying $M_{bc} > M_{B^0} - 3\sigma$ and
$|\Delta E| < 40 MeV$. 

\begin{figure}
\centerline{\mbox{\psfig{figure=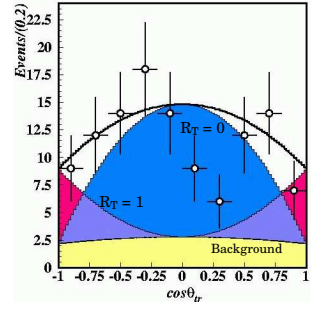,width=4cm}}}
\caption{Distribution of $\cos \theta_{tr}$ for $B^0 \rightarrow
D^{*+}D^{*-}$ candidates.}
\label{trans-dstar}
\end{figure}

The distribution of the transversity angle\cite{transversity} is 
measured in the
selected $B^0$ candidates as shown in Fig.~\ref{trans-dstar}.
The distribution can be expressed as a function of
the fraction of CP-odd component ($R_T$) as
\begin{small}
\begin{equation}
\frac{d\Gamma}{d\cos\theta_tr} = \frac{3}{4}(1-R_T)\sin^2\theta_{tr} + 
\frac{3}{2}R_T\cos^2\theta_{tr}.
\end{equation}
\end{small}
The predictions given by this formula for cases with $R_T = 0$ and $R_T = 1$ 
are also shown in the figure. As seen, the distribution of the data is 
close to the prediction with $R_T = 0$.

\subsection{$B^0\rightarrow D^{*-}\rho^{+}$}
The reconstruction of a $D^{*-}$ is performed in a similar manner as
described above. A $\rho^+$ is reconstructed from a pair of
$\pi^+$ and $\pi^0$ candidates. 
The $M_{bc}$ of a pair of reconstructed $D^*$ and $\rho$ is required
to be in the range 5.27 - 5.29 GeV/$c^2$ to and $\Delta E$
to satisfy $-100 < |\Delta E| < 50 MeV$. 

\begin{figure}
\centerline{\mbox{\psfig{figure=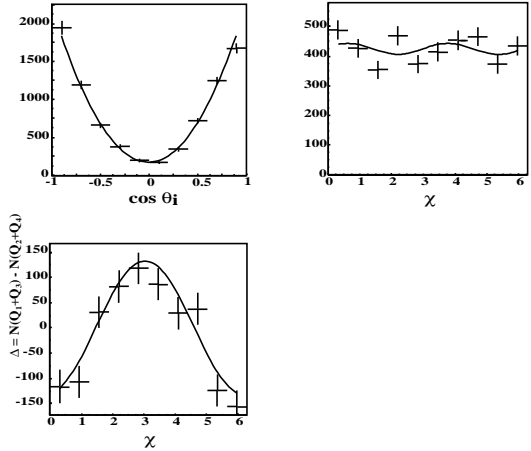,width=7cm}}}
\caption{Distributions of angles used in the angular analysis for
$B^0 \rightarrow D^{*-}\rho^+$ candidates. The distributions are
corrected for the acceptance and background contaminations.}
\label{ang_dstarrho}
\end{figure}

The time independent angular distribution is studied using the helicity
basis angles\cite{angleref}. 
The decay rates as a function of 
each of three projected angles (two helicity angles $\theta_i$ 
where $i=1,2$, and an angle between two decay planes
$\chi$) can be expressed as 
\begin{small}
\begin{gather}
\frac{d\Gamma}{d\cos \theta_i}  =  \frac{4\pi}{3}|A_0|^2\cos^2\theta_i 
 + \frac{2\pi}{3}(|A_{\perp}|^2+|A_{\parallel}|^2)\sin^2\theta_i, \\
\frac{d\Gamma}{d\chi}  = 
(\frac{4}{9}|A_0|^2+\frac{8}{9}|A_{\perp}|^2)\sin^2\chi \nonumber \\
~~~ +
(\frac{4}{9}|A_0|^2+\frac{8}{9}|A_{\parallel}|^2)\cos^2\chi
 + \frac{8}{9}Im(A_{\parallel}^*A_{\perp})\sin 2\chi
\end{gather}
\end{small}
where $A_0$, $A_{\parallel}$ and $A_{\perp}$ are the decay amplitudes in the
transversity basis. In addition, 
we also used $\chi$-differential decay rates
projected in 4 quadrants of a $\theta_1 - \theta_2$ plane
(angular regions from 0 to $\pi/2$ and $\pi/2$ to $\pi$ for each of
them) to observe
the effects in interference terms. 
The difference in the decay rates between two diagonally-summed 
quadrants ($\Delta$) can be written as
\begin{small}
\begin{equation}
\frac{d\Delta}{d\chi} =  -\frac{8}{9\sqrt{2}} Im(A_0^* A_{\perp}) \sin \chi
+ \frac{8}{9\sqrt{2}} Re(A_0^*A_{\parallel}) \cos\chi.
\end{equation}
\end{small}
Fig.~\ref{ang_dstarrho} shows the distributions of $\cos \theta_i$
($\theta_1$ and $\theta_2$ distributions are added), $\chi$, 
and $\Delta(\chi)$. The
measurement of the decay amplitudes is now in progress by fitting
these distributions with the theoretical formula given above
considering effects of the detector acceptance and background
contaminations. 

\section{Summary}
A full time dependent angular analysis is performed for $B^0\rightarrow 
J/\psi K^{*0}(K_S^0\pi^0)$ decays collected with the Belle detector at KEK
B-factory. The decay angles in the transversity basis ($\theta_{tr}$, 
$\phi_{tr}$ and $\theta_{K^*}$) and the decay time
difference of two $B$ mesons ($\Delta t$) are measured for each event
with the determination of the flavor of tag-side $B$ meson.
The values are
fitted to the theoretical distribution using an
unbinned maximum likelihood method considering the effects of 
detector acceptance, background contamination, $\Delta t$
resolution and wrong flavor tagging. 
From the fit, 
two CP violation parameters are determined to be
$\sin 2\phi_1  =  +0.13 \pm 0.51 \pm 0.06$, and 
$\cos 2\phi_1  =  +1.40 \pm 1.28 \pm 0.19$.
If we take the s-quark
helicity conservation choice of the amplitude phases and the
$\sin 2\phi_1$ value measured using other decay modes 
($0.72\pm0.07\pm0.04$)\cite{newcp}, the obtained
sign of $\cos 2\phi_1$ suggests the $2\phi_1$ to be in the range 
$0^{\circ} < 2\phi_1 < 90^{\circ}$ and
the other choice of $2\phi_1 \geq 90^{\circ}$ is not preferred
although the current statistics is not enough to conclude this.

A time independent angular analysis for other decay modes, 
$B^0 \rightarrow D^{*+}D^{*-}$, $B^0 \rightarrow D^{*-}\rho^+$, and 
$B^+ \rightarrow \rho^+ \rho^0$, are now being performed.


\end{document}